\documentclass[pre,twocolumn,amsmath,amssymb,nofootinbib,floatfix,superscriptaddress]{revtex4}

\usepackage{graphicx,bm}
\usepackage[normalem]{ulem} 

\makeatletter
\def\graphicscale{\twocolumn@sw{0.3}{0.4}}
\def\graphicthreescale{\twocolumn@sw{0.3}{0.4}}

\begin{document}

\title{Lattice gauge theories in the presence of a linear
  gauge-symmetry breaking}

\author{Claudio Bonati} 
\affiliation{Dipartimento di Fisica dell'Universit\`a di Pisa
        and INFN Largo Pontecorvo 3, I-56127 Pisa, Italy}

\author{Andrea Pelissetto}
\affiliation{Dipartimento di Fisica dell'Universit\`a di Roma Sapienza
        and INFN Sezione di Roma I, I-00185 Roma, Italy}

\author{Ettore Vicari} 
\affiliation{Dipartimento di Fisica dell'Universit\`a di Pisa
        and INFN Largo Pontecorvo 3, I-56127 Pisa, Italy}

\date{\today}

\begin{abstract}
  We study the effects of gauge-symmetry breaking (GSB) perturbations in
  three-dimensional lattice gauge theories with scalar fields. 
  We study this issue at transitions in which gauge
  correlations are not critical and the gauge symmetry only selects the
  gauge-invariant scalar degrees of freedom that become critical.
  A paradigmatic model in which this behavior is realized is 
  the lattice CP$^1$ model or, more generally, 
  the lattice Abelian-Higgs model with two-component complex
  scalar fields  and compact gauge fields. We consider this model 
  in the presence of a linear GSB perturbation. 
  The gauge symmetry turns out to be quite robust with
  respect to the GSB perturbation: the continuum limit is gauge-invariant also 
  in the presence of a finite small GSB term.  We also determine the phase 
  diagram of the model. It has one disordered phase
  and two phases that are tensor  and vector ordered, respectively.
  They are separated by continuous transition lines, which belong to the O(3),
  O(4), and O(2) vector universality classes, and which meet at 
  a multicritical point.
  We remark that the behavior at the CP$^1$ 
  gauge-symmetric critical point substantially differs from that 
  at transitions in which gauge correlations become critical,
  for instance at transitions in the noncompact lattice 
  Abelian-Higgs model that are controlled by the charged fixed point: 
  in this case the behavior is extremely sensitive to GSB perturbations.
 \end{abstract}

\maketitle

\section{Introduction}
\label{intro}

Gauge symmetries have been crucial in the development of 
theoretical models of fundamental
interactions~\cite{Wilson-74,Weinberg-book,ZJ-book}. They
also play a relevant role in statistical and 
condensed-matter physics
~\cite{Wen-book, Anderson-15, GASVW-18, Sachdev-19,
SSST-19, GSF-19, Wetterich-17, FNN-80, INT-80}. For instance,
it has been suggested that they 
may effectively emerge as low-energy effective
symmetries of many-body systems. However, because of the 
presence of microscopic gauge-symmetry
violations, it is crucial to understand the role 
of the gauge-symmetry breaking (GSB) perturbations. One would like 
to understand whether these perturbations do not change the low-energy 
dynamics---in this case the gauge-symmetric theory would describe the 
asymptotic dynamics even in the presence of some (possibly small) 
violations---or whether even small perturbations can destabilize the 
emergent gauge model--- consequently, a gauge-invariant dynamics would be 
observed only if an appropriate tuning of the model parameters is performed. 
This issue is also crucial in the context of analog
quantum simulations, for example, when controllable atomic systems are
engineered to effectively reproduce the dynamics of gauge-symmetric theoretical
models, with the purpose of obtaining physical information from the
experimental study of their quantum dynamics in laboratory.  Several proposals
of artificial gauge-symmetry realizations have been reported, see, e.g., the
review articles~\cite{ZCR-15,Banuls-etal-20,BC-20} and references therein, in
which the gauge symmetry is expected to effectively emerge in the low-energy
dynamics.

Previous studies of the critical behavior (or continuum limit) of
three-dimensional (3D) lattice gauge theories with scalar matter have shown the
emergence of two different scenarios. One possibility is that 
scalar-matter and gauge-field  correlations 
are both critical at the transition point. This
behavior is related to the existence of a charged fixed point (FP) in the
renormalization-group (RG) flow of the corresponding continuum gauge field
theory \cite{ZJ-book}. 
This is realized in some of the transitions observed in the
3D lattice Abelian-Higgs (AH) model with noncompact gauge fields \cite{BPV-21}.
Alternatively, it is possible that only scalar-matter correlations 
are critical at the transition. The gauge variables do not display
long-range critical correlations, although their 
presence  is crucial to identify the gauge-invariant scalar-matter degrees
of freedom that develop the critical behavior. This typically occurs
in lattice AH models with compact gauge variables.

In Ref.~\cite{BPV-21-gsb} we studied GSB perturbations in the 3D lattice
AH model with noncompact gauge field, at transitions controlled by the 
charged FP of the RG flow of
3D electrodynamics with multicomponent charged scalar
fields~\cite{HLM-74,FH-96,IZMHS-19,ZJ-book,MZ-03} (also called AH field theory).
In that case a
photon-mass GSB term, however small, gives rise to a drastic departure from the
gauge-invariant continuum limit of the statistical lattice gauge theory,
driving the system towards a different critical behavior (continuum limit).
This is due to the fact that a photon-mass term qualitatively
changes the phase diagram of the noncompact lattice AH theory~\cite{BPV-21}.
The Coulomb phase is no longer present and therefore, the nature of the 
transition to the Higgs phase, previously controlled by the charged fixed
point, varies. One ends up with a new critical behavior with noncritical 
gauge fields.

At transitions where gauge fields are not critical, the role 
of gauge symmetries is more subtle, but still crucial to determine the 
continuum limit. Even if gauge correlations are not critical, 
gauge fields prevent non-gauge invariant correlators from acquiring 
nonvanishing
vacuum expectation values and therefore, from developing long-range order.
Therefore, gauge symmetries effectively reduce the number of degrees of
freedom of the matter-field critical modes.  The 
lattice CP$^{N-1}$ model~\cite{PV-19-CP,TIM-05,TIM-06} or, more
generally, the lattice AH model with compact gauge
fields~\cite{PV-19-AH3d}, where an $N$-component complex scalar field is
gauge-invariantly coupled to a compact U(1) gauge field associated with the
links of the lattice, have transitions of this type. 
In these models the role of the U(1) gauge symmetry is
that of hindering some scalar degrees of freedom, i.e.,
those related to a local phase, from becoming critical.
As a consequence, the critical behavior or
continuum limit is driven by the condensation of a gauge-invariant tensor
scalar operator, and the corresponding
continuum field theory is associated with a Landau-Ginzburg-Wilson (LGW) field
theory with a tensor field, but without gauge fields. 
For $N=2$, the LGW approach predicts that 
the continuum limit of the CP$^1$ model (or of the $N=2$ 
lattice compact AH model) is equivalently described by the 
LGW field theory for a three-component real  vector
field with O(3) global symmetry.

In this paper we investigate the effects of perturbations breaking the gauge
symmetry when gauge-field correlations are not critical and the gauge symmetry
acts only to prevent some of the matter degrees of freedom from becoming
critical.  For this purpose we consider a lattice AH model with compact gauge
fields, focusing on the case of two scalar components. We consider the most
natural and simplest GSB term, adding to the Hamiltonian (or action, in the
high-energy physics terminology) a term that is linear in the gauge link
variable.  This GSB term is similar to the photon mass term considered in
noncompact lattice AH models, and indeed it is equivalent to it in the limit
of small noncompact field $A_{{\bm x},\mu}$, as can be immediately seen by expanding the
exponential relation between the compact and  noncompact gauge fields,
i.e., $\lambda_{{\bm x},\mu}=\exp(iA_{{\bm x},\mu})$.  At
variance with what was obtained in Ref.~\cite{BPV-21-gsb}, where we considered
GSB perturbations at transitions controlled by charged FPs, in the present case
the linear GSB perturbation does not change the continuum limit, at least for a
sufficiently small finite strength of the perturbation.

\begin{figure}[tbp]
\includegraphics[width=0.95\columnwidth, clip]{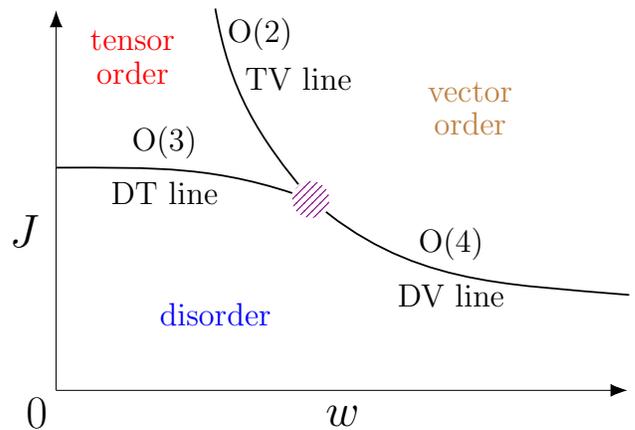}
\caption{Sketch of the phase diagram of the lattice CP$^1$ model in the
   presence of the GSB term 
   $H_b=-w\sum_{{\bm x},\mu} {\rm Re}\,\lambda_{{\bm x},\mu}$.
   This is a particular case of the lattice AH model with two-component
   scalar matter and compact gauge variables for $\gamma=0$. 
   The phase diagram is characterized by three
   different phases: a disordered phase (small $J$), a tensor-ordered 
   phase where the tensor operator $Q$ condenses (large $J$ and 
   small $w$), and a vector-ordered phase where the vector field 
   ${\bm z}_{\bm x}$ condenses (large $J$ and $w$). These phases are 
   separated by the DT, DV, and TV transition lines, where CP$^1$/O(3),
   O(4) vector, and O(2) vector critical behavior is observed.}
\label{phdiagn2}
\end{figure}

In Fig.~\ref{phdiagn2} we
anticipate a sketch of the phase diagram of the lattice CP$^1$ model in the
presence of a GSB term whose strength is controlled by the parameter $w$.
It presents three
different phases: a high-temperature (small $J$) 
disordered phase, and two low-temperature (large $J$)
phases with different orderings.  When the GSB perturbation is small, the
low-temperature phase 
is qualitatively analogous to that of the 
gauge-invariant model, i.e., it is characterized by the
condensation of a bilinear tensor scalar operator. In the 
RG language, the GSB perturbation turns out to be irrelevant at the FP of the
gauge-invariant theory. On the other hand, when the GSB parameter $w$ is 
sufficiently large, there is a second ordered phase characterized by the 
condensation of the vector scalar field. These phases are separated by  
three different transition lines, meeting at a multicritical point. 
Along the disordered-tensor (DT) transition line we observe the same 
critical behavior as in the CP$^1$ model: the tensor degrees of freedom 
behave as in the O(3) vector model. The GSB term is 
irrelevant and the continuum limit of the model with a finite GSB breaking
is the same as that of the gauge-invariant model. Along the DV line, we observe
instead a different continuum limit: vector and tensor degrees of freedom
behave as in the O(4) model. Finally, on the TV line we observe an
O(2) vector 
critical behavior.

The paper is organized as follows.  In Sec.~\ref{model} we define the
lattice AH model with compact gauge variables and the linear GSB
perturbation. In Sec.~\ref{obs} we define the observables that we
consider in the numerical study and review  the main properties of
the finite-size scaling (FSS) analyses employed.  In
Sec.~\ref{phdiacrbeh} we discuss some limiting cases where
the thermodynamic behavior is known and propose a possible phase diagram for 
the model.  In Sec.~\ref{numres} we discuss our
numerical results for $N=2$, which support the conjectured
phase diagram. Finally, in Sec.~\ref{conclu} we summarize and draw
our conclusions. In the Appendix we report some universal curves 
in the O($N$) vector model, that allow us to identify the 
universality class of the different transitions.

\section{The model}
\label{model}

\subsection{The lattice Abelian-Higgs model with compact gauge variables}
\label{coAH}

A compact lattice formulation of the three-dimensional AH model is obtained by
associating complex $N$-component unit vectors ${\bm z}_{\bm x}$ with the sites
${\bm x}$ of a cubic lattice, and U(1) variables $\lambda_{{\bm x},\mu}$ with
each link connecting the site ${\bm x}$ with the site ${\bm x}+\hat\mu$ (where
$\hat\mu=\hat{1},\hat{2},\ldots$ are unit vectors along the positive lattice
directions).  The partition function of the system reads
\begin{eqnarray}
&&Z = \sum_{\{{\bm z},\lambda\}} e^{-H_{\rm AH}({\bm z},\lambda)}\,,
\label{partfunc}\\
&& H_{\rm AH}({\bm z},\lambda) = H_z({\bm z},\lambda) + H_{\lambda}(\lambda)\,
.
\label{Hdef}
\end{eqnarray}
We define
\begin{eqnarray}
H_z = - J\, N
\sum_{{\bm x}, \mu}
2\,{\rm Re}\, \lambda_{{\bm x},\mu}\,  \bar{\bm z}_{\bm x} 
\cdot  {\bm z}_{{\bm x}+\hat\mu} \,,
\label{Hzdef}
\end{eqnarray}
where the sum runs over all lattice links, and
\begin{equation}
H_\lambda = - \gamma \sum_{{\bm x},\mu>\nu} 
2\,{\rm Re}\, \Pi_{{\bm x},{\mu}\nu} \, ,
\label{Hladef}
\end{equation}
where \
\begin{equation}
\Pi_{{\bm x},{\mu}\nu} = 
\lambda_{{\bm x},{\mu}} \,\lambda_{{\bm x}+\hat{\mu},{\nu}} 
\,\bar{\lambda}_{{\bm x}+\hat{\nu},{\mu}}  
  \,\bar{\lambda}_{{\bm x},{\nu}} \,,
\end{equation}
and the sum runs over all plaquettes of the cubic lattice.  The AH
Hamiltonian is invariant under the global SU($N$) transformations
\begin{equation}
z_{\bm x} \to U z_{\bm x}\,,\qquad U\in\mathrm{SU}(N)\,,
\label{sunsym}
\end{equation}
and the local U(1) gauge transformations
\begin{eqnarray}
z_{\bm x} \to e^{i\theta_{\bm x}}  z_{\bm x}\,,\quad
\lambda_{{\bm x},\mu} \to e^{i\theta_{\bm x}}  \lambda_{{\bm x},\mu}
e^{-i\theta_{{\bm x}+\hat{\mu}}}\,,
\label{gaugsym}
\end{eqnarray}
where $\theta_{\bm x}$ is an arbitrary space-dependent real function.
The parameter $\gamma\ge 0$ plays the role of inverse gauge coupling.

For $\gamma=0$, the model is 
a particular lattice formulation of the 3D CP$^{N-1}$ model,
which is quadratic in the scalar-field variables and linear in
the gauge variables. We can obtain a lattice formulation
without explicit gauge fields by integrating 
out the link variables. We obtain
\begin{equation}
Z = \sum_{\{{\bm z},\lambda\}} e^{-H_z({\bm z},\lambda)} 
= \sum_{\{{\bm z}\}}
\prod_{{\bm x},\mu} 
I_0\left(2 J N |\bar{\bm z}_{\bm x} \cdot {\bm z}_{{\bm
    x}+\hat\mu}|\right)\,,
\label{partzint}
\end{equation}
where $I_0(x)$ is a modified Bessel function.  The corresponding
effective Hamiltonian is
\begin{equation}
H_{{\rm eff}} = - \sum_{{\bm x},\mu}  
\ln I_0\left(2 J N |\bar{\bm z}_{\bm x} \cdot
    {\bm z}_{{\bm x}+\hat\mu}|\right)\,,
\label{hlaeff}
\end{equation}
which is invariant under the gauge transformations (\ref{gaugsym}) even
in the absence of gauge fields. For $J$ small, since  
$I_0(x) = 1 + x^2/4 + O(x^4)$,
the Hamiltonian 
$H_{{\rm eff}}$ simplifies to 
\begin{equation}
H_{CP} = - J^2 N^2 \sum_{{\bm x},\mu}   |\bar{\bm z}_{\bm x} \cdot
    {\bm z}_{{\bm x}+\hat\mu}|^2,
\end{equation}
which represents another equivalent formulation of the CP$^{N-1}$ model.

The compact AH model presents two phases when varying $J$ and
$\gamma$ \cite{PV-19-AH3d}, separated by a transition line, whose 
nature does not depend on the gauge parameter $\gamma$.
The appropriate order parameter is the bilinear gauge-invariant
operator
\begin{equation}
Q_{{\bm x}}^{ab} = \bar{z}_{\bm x}^a z_{\bm x}^b - {1\over N}
\delta^{ab}\,,
\label{qdef}
\end{equation}
which is a hermitian and traceless $N\times N$ matrix.  It transforms
as $Q_{{\bm x}} \to {U}^\dagger Q_{{\bm x}} \,{U}$ under the global
SU($N$) transformations. Its condensation signals the
spontaneous breaking of the global SU($N$) symmetry.

\subsection{Linear breaking of the U(1) gauge symmery}
\label{u1gb}

In the following we investigate the
effects of perturbations breaking gauge invariance. In particular, 
we consider the simplest gauge-breaking perturbation that is 
linear in the $\lambda$ variables.  We consider the 
extendend Hamiltonian 
\begin{equation}
H_e({\bm z},\lambda) = H_{\rm AH}({\bm z},\lambda) + H_{b}(\lambda)\, ,
\label{Hext}
\end{equation}
where
\begin{equation}
H_{b} = - w \sum_{{\bm x},\mu} {\rm Re} \,\lambda_{{\bm x},\mu}\,.
\label{linpert}
\end{equation}
Note that, if we perform the change of variables
\begin{equation}
{\bm z}_{\bm x} \to (-1)^{x_1+x_2+x_3}{\bm z}_{\bm x}\ ,\quad
\lambda_{{\bm x},\mu}\to -\lambda_{{\bm x},\mu}\,,
\label{F-to-AF}
\end{equation}
we reobtain the action (\ref{Hext}), with $w$ replaced by $-w$.
Thus, the phase diagram is independent of the sign of $w$ (but, for 
$w < 0$, the relevant vector order parameters would be staggered quantities).
Thus, in the following we only consider the case $w\ge 0$.  

When $\gamma=0$, one can straightforwardly integrate out the link variables
$\lambda$, obtaining
\begin{eqnarray}
Z &=& \sum_{\{{\bm z},\lambda\}} e^{-H_{z}({\bm z},\lambda) -
  H_{b}(\lambda)} \nonumber\\
  &=& \sum_{\{{\bm z}\}} \prod_{{\bm x},\mu} I_0\left(2 J
  N |\hat{w}+\bar{{\bm z}}_{\bm x} \cdot {\bm z}_{{\bm x}+\hat\mu}|\right),
\label{partzint2}
\end{eqnarray}
where $\hat{w}=w/(2JN)$.  The corresponding effective Hamiltonian
reads
\begin{equation}
H_{{\rm eff}} = - 
\sum_{{\bm x},\mu}  
\ln I_0\left(2 J N |\hat{w} + \bar{\bm z}_{\bm x} \cdot 
{\bm z}_{{\bm x}+\hat\mu}|\right)\,.
\label{hlaeff2}
\end{equation}
The Bessel function should be irrelevant for the critical behavior. 
Since, the argument of the Bessel function can be equivalently written as 
\begin{equation}
|\hat{w} + \bar{\bm z}_{\bm x} \cdot {\bm z}_{{\bm x}+\hat\mu}|^2 
 = \hat{w}^2  + 
2 \hat{w}\, \hbox{Re}\, (\bar{\bm z}_{\bm x} \cdot {\bm z}_{{\bm x}+\hat\mu}) + 
|\bar{\bm z}_{\bm x} \cdot {\bm z}_{{\bm x}+\hat\mu}|^2,
\end{equation}
we expect that, for infinite gauge coupling, the model has the same 
critical behavior as a model with Hamiltonian
\begin{equation}
 \widetilde{H}_{{\rm eff}} = 
-  J^2 N^2 \sum_{{\bm x},\mu}\Big[
|\bar{\bm z}_{\bm x} \cdot {\bm z}_{{\bm x}+\hat\mu}|^2 + 
2 \hat{w} \,{\rm Re} \,\bar{\bm z}_{\bm x} \cdot {\bm z}_{{\bm x}+\hat\mu}
\Big]\,.
\label{hlaeffexpmix}
\end{equation}
Such a Hamiltonian is the sum of a U($N$) invariant CP$^{N-1}$ model and 
an O($2N$) invariant vector model, which represents now the gauge-breaking 
perturbation. Thus, the introduction of the linear gauge-breaking term 
(\ref{linpert}) is equivalent to the addition of a ferromagnetic vector
interaction ${\rm Re}(\bar{\bm z}_{\bm x} \cdot {\bm z}_{{\bm x}+\hat\mu})$ 
among 
the vector fields. Although
this has been shown for $\gamma = 0$, we expect this equivalence  to hold 
at criticality for any positive finite $\gamma$.

We would like now to show that, at transitions where the vector fields show a
critical behavior, one observes an enlarged O($2N$) symmetry. 
For this purpose, note that the CP$^{N-1}$ interaction 
in Eq.~(\ref{hlaeffexpmix}) can be rewritten as 
\begin{equation}
|\bar{\bm z}_{\bm x} \cdot {\bm z}_{{\bm x}+\hat\mu}|^2 = 
({\rm Re} \,\bar{\bm z}_{\bm x} \cdot {\bm z}_{{\bm x}+\hat\mu})^2 +
({\rm Im}\, \bar{\bm z}_{\bm x} \cdot \Delta_\mu {\bm z}_{\bm x})^2, 
\label{hlaeffexp}
\end{equation}
where $\Delta_\mu {\bm z}_{\bm x} \equiv {\bm z}_{{\bm x}+\hat{\mu}} - {\bm
z}_{\bm x}$, and we used the fact that $\bar{\bm z}_x \cdot {\bm z}_x=1$.
The first term represents an additional O($2N$) invariant 
vector ferromagnetic interaction, while the second one 
is only invariant under
U($N$) transformations. We will now argue that the latter term is
irrelevant at the O($2N$) fixed point. For this
purpose, we consider the usual LGW approach and define a 
real field $\Psi_{ai}$,  $a=1,...,N$ and $i=1,2$, which represents 
a coarse-grained version of the field ${\bm z}_{\bm x}$, the 
correspondence being
\begin{equation}
{\rm Re} \, {\bm z}_{\bm x}^a \to \Psi_{a1}({\bm x}) \,,\qquad
{\rm Im} \, {\bm z}_{\bm x}^a \to \Psi_{a2}({\bm x}) \,,
\label{psicorr}
\end{equation}
The coarse-grained Hamiltonian corresponding to model (\ref{hlaeffexpmix}) 
is given by
\begin{eqnarray}
&&L_{\rm LGW} = {1\over 2} \sum_{ai,\mu} (\partial_\mu  \Psi_{ai})^2
+ {r\over 2} \sum_{ai} \Psi_{ai}^2 \label{lgwrfixed}\\
&&\;\; +
{u\over 4!} (\sum_{ai} \Psi_{ai}^2)^2 
+ v \sum_{a,\mu} (\Psi_{a1}\,\partial_\mu \Psi_{a2} -
\Psi_{a2}\,\partial_\mu \Psi_{a1})^2\,.
\nonumber
\end{eqnarray}
The first three terms are O($2N$) symmetric and represent the 
coarse-grained version of the O($2N$)-invariant part of the Hamiltonian.
The last term corresponds to the O($2N$)-symmetry breaking
term, i.e., to the last term in Eq.~(\ref{hlaeffexp}).  Since this
quartic term contains two
derivatives, its naive dimension is six close to four
dimensions and, therefore, it is generally expected to be irrelevant at the
three-dimensional O($2N$)-symmetric fixed point.  This symmetry
enlargement is a general result that holds for generic 3D vector
systems with global U($N$) invariance (without gauge
symmetries)~\cite{BPV-21-gsb}. It should be stressed that 
this symmetry enlargement is only present in the large-scale 
critical behavior. Moreover, it  assumes that the vector fields are critical 
at the transition, since the LGW theory is defined in terms of 
coarse-grained $z$ fields. Therefore, no O($2N$) behavior is expected 
at transitions where vector correlations are short-ranged (as we discuss 
below this may also occur for finite values of $w$).

\subsection{Axial-gauge fixing}

In a lattice gauge theory the redundancy arising from the gauge
symmetry can be eliminated by adding a gauge fixing, which leaves 
gauge-invariant observables unchanged. For example,
one  may consider the {\em axial gauge},
obtained by fixing the link variable $\lambda_{{\bm x},\mu}$ along one
direction,
\begin{equation}
{\rm axial}\;{\rm gauge}: \qquad \lambda_{{\bm x},3}=1\,.
\label{axialgauge}
\end{equation}
For periodic boundary conditions, this constraint cannot be satisfied on
all sites, and a complete gauge fixing is obtained by  setting
$\lambda_{{\bm x},\mu}=1$ on a maximal lattice
tree~\cite{Creutz-book,ItzDro-book}, that necessarily involves some links 
that belong to an orthogonal plane. To avoid this problem, one can use
$C^*$ boundary conditions~\cite{KW-91,LPRT-16,BPV-21-gsb}. In this case, 
one can require condition (\ref{axialgauge}) on all sites $\bm x$. 

In the following we will study the phase diagram of the model (\ref{Hext})
in the presence of the axial gauge fixing. Note that, also in 
the case, the phase diagram is independent of the sign of $w$. Instead 
of Eq.~(\ref{F-to-AF}) one should consider the mapping 
${\bm z}_{\bm x} \to (-1)^{x_1+x_2}{\bm z}_{\bm x}$, $\lambda_{{\bm x},\mu} 
\to - \lambda_{{\bm x},\mu}$.

It is interesting 
to write down the effective Hamiltonian that is obtained by integrating the 
gauge fields. In the axial gauge, repeating the same arguments presented in 
Sec.\ref{u1gb}, we obtain 
\begin{eqnarray}
 \widetilde{H}_{{\rm eff}} &= &
-  J^2 N^2 \sum_{{\bm x},\mu}\Big[
|\bar{{\bm z}}_{\bm x} \cdot {\bm z}_{{\bm x}+\hat\mu}|^2 + 
2 \hat{w} \,{\rm Re} \,\bar{\bm z}_{\bm x} \cdot {\bm z}_{{\bm x}+\hat\mu}
\Big]\, \nonumber \\
&& -2 J N \sum_{{\bm x} }
 {\rm Re} \,\bar{\bm z}_{\bm x} \cdot {\bm z}_{{\bm x}+\hat{3}},
\label{Haxial-eff}
\end{eqnarray}
where, in the first sum, $\mu$ takes only the values 1 and 2. We this obtain 
a stacked CP$^{N-1}$-O($2N$) ferromagnetic system, 
in which the different layers are
ferromagnatically coupled by a standard vector interaction.

\section{Observables and FSS analyses}
\label{obs}


In our numerical analysis we consider the vector and tensor two-point
functions of the scalar field, respectively defined as
\begin{equation}
  G_V({\bm x},{\bm y}) =
  \hbox{Re}\ \langle \bar{\bm z}_{\bm x} \cdot {\bm z}_{\bm y}\rangle
  \label{vectorfu}
\end{equation}
and
\begin{equation}
G_T({\bm x},{\bm y}) = \langle {\rm Tr}\, Q_{\bm x}Q_{\bm y} \rangle\,,  
\label{gxyp}
\end{equation}
where $Q$ is the bilinear operator defined in Eq.~(\ref{qdef}).
The corresponding susceptibility and second-moment correlation length
are defined by the relations
\begin{eqnarray}
&&\chi_{V/T} =  \sum_{{\bm x}} G_{V/T}({\bm x}) = 
\widetilde{G}_{V/T}({\bm 0}), 
\label{chisusc}\\
&&\xi_{V/T}^2 \equiv  {1\over 4 \sin^2 (\pi/L)}
{\widetilde{G}_{V/T}({\bm 0}) - \widetilde{G}_{V/T}({\bm p}_m)\over 
\widetilde{G}_{V/T}({\bm p}_m)},
\label{xidefpb}
\end{eqnarray}
where $\widetilde{G}_{V/T}({\bm p})=\sum_{{\bm x}} e^{i{\bm p}\cdot
  {\bm x}} G_{V/T}({\bm x})$ is the Fourier transform of $G_{V/T}({\bm
  x})$, and ${\bm p}_m = (2\pi/L,0,0)$.  In our FSS analysis we
consider the ratios
\begin{equation}
  R_{V/T} = \xi_{V/T}/L\,
  \label{rvt}
\end{equation}
associated with the vector and tensor correlation lengths, and the
vector and tensor Binder parameters
\begin{eqnarray}
&&  U_{V/T} = \frac{\langle \mu_{V/T}^2\rangle}{\langle \mu_{V/T} \rangle^2} \,,
  \label{binderdef}\\
 &&\mu_{V} = \sum_{{\bm x},{\bm y}} {\rm Re}\,\bar{\bm z}_{\bm x}\cdot {\bm z}_{\bm y}\,,
  \qquad
  \mu_{T} = \sum_{{\bm x},{\bm y}} {\rm Tr}\, Q_{\bm x}Q_{\bm y}\,.
  \nonumber
\end{eqnarray}  
We recall that, at continuous phase transitions driven by a relevant
Hamiltonian parameter $r$, the RG invariant quantities associated with
the critical modes are expected to scale as (we denote a
generic RG invariant quantity by $R$)~\cite{PV-02}
\begin{eqnarray}
R(L,w) &\approx & f_R(X) \,,\quad X =(r-r_c)\,L^{1/\nu}\,,
 \label{scalbeh}
\end{eqnarray}
where $\nu$ is the critical exponent associated with the correlation
length.  Scaling corrections decaying as $L^{-\omega}$ have been
neglected in Eq.~(\ref{scalbeh}), where $\omega$ is the exponent
associated with the leading irrelevant operator. The function $f_R(X)$
is universal up to a multiplicative rescaling of its argument.  In
particular, $U^*\equiv f_U(0)$ and $R_\xi^*\equiv f_{R_\xi}(0)$ are
universal, depending only on the boundary conditions and the aspect
ratio of the lattice. To verify universality, we will 
plot $U$ versus $R_\xi$. The data are expected to approach a universal
curve, i.e., 
\begin{equation}
U = F_U(R) + O(L^{-\omega})\,,
\label{yfur}
\end{equation}
where $F_U(R)$ is universal and independent of any normalization. It only
depends on the boundary conditions and aspect ratio of the lattice.

\section{Phase diagram and critical behaviors}
\label{phdiacrbeh}

Before presenting the results of the numerical 
simulations, we discuss some limiting cases that 
allow us to determine the phase diagram of the model. 
Because of the symmetry under the change of the sign of $w$, 
we only consider the case $w \ge 0$.

\subsection{The model for $w=0$ and $w=\infty$} 
\label{sec4.A}

For $w=0$ we recover the lattice AH model with
compact U(1) gauge variables. Its phase diagram has been extensively studied in
the literature, see Refs.~\cite{PV-19-AH3d,PV-20}. For 
any $\gamma\ge 0$, 
there is a disordered phase for small $J$ and an ordered phase for large
$J$, where the gauge-invariant operator $Q$
defined in  Eq.~(\ref{qdef}) condenses. For $N=2$ 
the transition occurs at 
~\cite{PV-19-AH3d} $J_c=0.7102(1)$ for $\gamma=0$,
$J_c=0.4145(5)$ for $\gamma=0.5$, and $J_c=0.276(1)$ for $\gamma=1$. 
For this value of $N$ the transitions are
continuous and belong to the O(3) vector universality
class~\cite{PV-02} as the transition in the CP$^1$ model corresponding to 
$\gamma = 0$.  Accurate estimates of the O(3) critical exponents can be
found in Refs.~\cite{Hasenbusch-20, Chester-etal-20-o3, KP-17, HV-11, PV-02,
CHPRV-02, GZ-98}. In the following we use the estimate of the 
correlation-length exponent $\nu$ of Ref.~\cite{Hasenbusch-20}, 
$\nu= 0.71164(10)$.  
For $N > 2$, transitions are instead of first order \cite{PV-19-AH3d,PV-20}.

In the limit $\gamma\to\infty$, the plaquette operator 
$\Pi_{{\bm x},{\mu}\nu}$ converges to 1. In infinite volume,   
this implies that $\lambda_{{\bm x},\mu} = 1$ modulo 
gauge transformations. We thus recover the O$(2N)$ vector model.
For $N=2$ the relevant model is the O(4) model, which has a transition 
at~\cite{BFMM-96,CPRV-96,BC-97} $J_c=0.233965(2)$. 
Estimates of the O(4) critical exponents can be
found in Refs.~\cite{HV-11,PV-02}: for example, $\nu=0.750(2)$.
The O(4) fixed point at $\gamma=\infty$
is unstable with respect to nonzero gauge couplings. For finite values of 
$\gamma$ it only gives rise to crossover phenomena~\cite{PV-19-AH3d}.

It is important to stress that, in the limit $\gamma\to \infty$, one 
obtains O($2N$) behavior only for gauge-invariant quantities, as 
$\lambda_{{\bm x},\mu} = 1$ modulo gauge transformations. For
instance, tensor correlations in the AH model converge 
to the corresponding O($2N$) correlations in the limit. 
Instead, quantities that are not gauge
invariant are not related to the corresponding quantities of the 
O($2N$) model. For instance, vector correlations satisfy 
$G_V({\bm x}) = \delta_{{\bm x},{\bm 0}}$ for any $\gamma$ and are 
therefore not related to vector correlations in the O($2N$) model.

In a finite volume with periodic boundary conditions, 
the limit $\gamma\to \infty$ is more subtle. Indeed,
Polyakov loops, i.e.,
the product of the gauge fields along nontrivial paths that wrap around the 
lattice, do not order in the limit. This implies that one cannot set 
$\lambda_{{\bm x},\mu} = 1$ on all sites. Rather, on some boundary 
links one should set 
\begin{eqnarray}
&& \lambda_{(L,n_2,n_3),1} = \tau_1,  \nonumber \\
&& \lambda_{(n_1,L,n_3),2} = \tau_2,  \nonumber \\
&& \lambda_{(n_1,n_2,L),3} = \tau_3, 
\end{eqnarray}
where $\tau_1$, $\tau_2$, and 
$\tau_3$ are three space-independent boundary 
phases that should be integrated over. We thus obtain an O($2N$) model 
with U(1)-fluctuating boundary conditions. This argument 
generalizes a similar result that holds in systems with real
fields and ${\mathbb Z}_2$ gauge invariance \cite{Hasenbusch-96}. In the latter
case, one obtains models with fluctuating periodic-antiperiodic boundary 
conditions. 

The behavior for $w\to \infty$ is definitely simpler.
In this limit 
$\lambda_{{\bm x},\mu}$ converges to 1 trivially. Thus, for any $\gamma$
both gauge-invariant and non gauge-invariant quantities behave as in 
the standard O($2N$) vector model. 

\subsection{The model for $J\to \infty$}
\label{sec4.B}

For $J\to\infty$, the relevant configurations are those that minimize the 
Hamiltonian term $H_z$. This implies that 
\begin{equation} 
    {\bm z}_{\bm x} = \lambda_{{\bm x},\mu} {\bm z}_{{\bm x} + \hat{\mu}}.
\label{zJinfinity}
\end{equation}
By repeated application of this relation, one can verify that 
the product of the gauge fields along any lattice loop, including nontrivial
loops that wrap around the lattice, is always 1.
Therefore, the gauge variables can be written (in a finite volume with 
periodic boundary conditions, too) as
\begin{equation}
  \lambda_{{\bm x},\mu} = \bar\psi_{\bm x} \psi_{{\bm x}+\hat\mu}\,,
  \qquad \psi_{\bm x} \in {\rm U}(1) \,.
\label{puregauge}
\end{equation}
Substituting in  Eq.~(\ref{zJinfinity}), we obtain
\begin{equation}
  {\bm z}_{\bm x} \psi_{\bm x} = {\bm z}_{{\bm x} + \hat{\mu}} 
    \psi_{{\bm x} + \hat{\mu}}.
\end{equation}
We can thus define a constant unit-length
vector ${\bm v} = {\bm z}_{\bm x} \psi_{\bm x}$, so that
\begin{equation}
{\bm z}_{\bm x} = \bar{\psi}_{\bm x} \, {\bm v}.
\label{puregaugez}
\end{equation}
Because of Eqs.~(\ref{puregauge}) and (\ref{puregaugez}), 
the only nontrivial Hamiltonian term 
is $H_b$ defined in Eq.~(\ref{linpert}), which reduces to 
the  XY Hamiltonian
\begin{equation}
H_b = - w \sum_{{\bm x},\mu} {\rm Re}\, \bar{\psi}_{\bm x} \,
  \psi_{{\bm x}+\hat\mu} \,.
\label{hbJinf}
\end{equation}
The XY model 
has a continuous transition at~\cite{CHPV-06,DBN-05} $w_c= 0.4541652(11)$.
Estimates of the critical exponents can be
found in Refs.~\cite{CHPV-06,Hasenbusch-19,CLLPSSV-19}; for example,
$\nu=0.6717(1)$. Thus, for $J=\infty$ 
we expect two phases, which can be equivalently characterized by using 
correlations of the gauge field or vector correlations.
Indeed, relation (\ref{puregaugez}) implies 
\begin{equation}
  G_V({\bm x},{\bm y}) = \hbox{\rm Re}\,
  \langle \bar{\bm z}_{\bm x} \cdot {\bm z}_{\bm y}\rangle = 
  \hbox{\rm Re}\, \langle \bar{\psi}_{\bm x} \, \psi_{\bm y}\rangle \,.
  \label{gvinfJ}
  \end{equation}

\subsection{The model for $\gamma=\infty$}
\label{sec4.C}

We have already discussed this limit for $w=0$. In that case,
gauge-invariant quantities behave as in  
the O($2N$) model, although with different boundary conditions. 
For $w\not=0$, as the model is not gauge invariant, we cannot 
get rid of the gauge fields. However, we can still rewrite the gauge fields 
as in Eq.~(\ref{puregauge})---for the moment, we ignore the subtleties 
of the boundary conditions---and therefore obtain the following 
effective Hamiltonian
\begin{eqnarray}
H_{\gamma=\infty} &=& 2 J N
\sum_{{\bm x}, \mu}
\hbox{\rm Re}\, 
\bar{\psi}_{\bm x} \psi_{{\bm x}+\hat\mu} \,  \bar{\bm z}_{\bm x}
\cdot  {\bm z}_{{\bm x}+\hat\mu} 
\nonumber \\
&& 
- w \sum_{{\bm x},\mu} {\rm Re} \, \bar{\psi}_{\bm x} \psi_{{\bm x}+\hat\mu}.
\end{eqnarray}
If we define new variables ${\bm Z}_{\bm x} = {\bm z}_{\bm x} \psi_{\bm x}$, 
the Hamiltonian $H_{\gamma=\infty}$ is the sum of two contributions 
that corresponding to two independent models: an O($2N$) model with 
coupling $J$ and fields ${\bm Z}_{\bm x}$ and an XY model with coupling 
$w$ and fields $\psi_{\bm x}$. Thus, four different phases appear,
that are separated by the critical lines $w=w_c$ and $J=J_c$  meeting at 
a tetracritical point, with O($2N$) and XY decoupled multicriticality.  
Correlation functions of the bilinear fields $Q_{\bm x}$ and 
$\lambda_{{\bm x},\mu}$ are only sensitive to  the O($2N$) and XY critical 
behavior, respectively. Vector correlations instead are sensitive to 
both transitions, since 
\begin{equation}
G_V({\bm x}) = \hbox{Re}\, 
   \langle \bar{\bm Z}_{\bm 0} \cdot {\bm Z}_{\bm x}\rangle 
   \langle {\psi}_{\bm 0} \bar{\psi}_{\bm x}\rangle.
\end{equation}
In particular, the vector fields only order for $w > w_c$ and $J > J_c$, 
where both ${\bm Z}_{\bm x}$ and $ \psi_{\bm x}$ show long-range 
correlations.

The decoupling of the XY and O($2N$) degrees of freedom only occurs 
in infinite volume. For finite systems with periodic boundary conditions, 
for $\gamma\to \infty$ one obtains U(1) fluctuating boundary conditions 
for both fields, with the same boundary fields $\tau_\mu$. In this 
case, a complete decoupling is not realized.

\subsection{The model for $J=0$} 
\label{sec4.D}

For $J=0$ we obtain a pure U(1) gauge theory in the presence of the 
gauge-breaking term (\ref{linpert}). For $\gamma = \infty$, as discussed 
in Sec.~\ref{sec4.C}, an XY transition occurs at $w = w_c$, where the 
gauge field 
$\lambda_{{\bm x},\mu}$ becomes critical. As we discuss below, this 
transition disappears for finite values of $\gamma$. For large values of
$\gamma$ only a crossover occurs for $w\approx w_c$.

\subsection{The phase diagram} \label{sec4.E}

The limiting cases we have discussed  above and the numerical results
that we present below allow us to conjecture the phase diagram of the model. 
In Fig.~\ref{phdiagn2} we sketch the  $J$-$w$ phase diagram for
$\gamma=0$ and $N=2$. It is supported by the numerical results 
and is consistent with the limiting cases reported 
above. We expect three different phases. For small values of $J$ there is 
a disordered phase, while for large $J$ there are two different ordered
phases.
For  small $w$ and large $J$,  the tensor operator $Q$ condenses,
while the vector correlation $G_V$, defined in Eq.~(\ref{vectorfu}),
is short ranged. In this phase the model behaves as for $w=0$: the
gauge-symmetry breaking is irrelevant and gauge invariance is recovered in 
the critical limit. On the other hand, for large $w$, both  vector and 
tensor correlations are long-ranged. In the latter phase, the ordered behavior 
of tensor correlations is just a consequence of the ordering of the 
vector variables ${\bm z}_{\bm x}$. We do not expect 
$\gamma$ to be relevant, and therefore, we expect the same phase diagram 
for any finite $\gamma$.

Note that, for $\gamma=\infty$, the 
multicritical point is tetracritical and four lines are present. In 
particular, there is a line where $\lambda_{{\bm x},\mu}$  orders,
while both vector and tensor correlations are short-ranged. 
We have no evidence of this transition line for finite values of $\gamma$,
at least for $N=2$, the only case we consider. Simulations for $J=0$ and 
for small values of $J$, in the
parameter region where vector and tensor correlations are both disordered,
observe crossover effects but no transitions. 

The three phases mentioned above are separated by three transition lines, 
whose nature depends on the value of $N$. For $N=2$, transitions are 
continuous. Along the DT line, which separates the disordered phase from the
tensor-ordered and vector-disordered phase, we expect the transition to 
belong to the same universality class as the transition for
$w=0$. Thus, the transitions should belong to the O(3) vector universality 
class.  Along the DV line, which separates the disordered phase
from the vector-ordered and tensor-ordered phase, we expect the transition to 
belong to the same universality class as the transition for
$w=\infty$. It should belong to the O(4) vector universality class,
consistently with the LGW argument presented in Sec.~\ref{u1gb}.
Finally, along the TV line that starts at $J=\infty$ and separates
the two tensor-ordered phases, 
the critical behavior should be associated with the phases of the scalar
variables. Therefore, the most
natural hypothesis is that transitions  belong to the O(2) or XY universality
class, as it occurs for $J\to\infty$. 
The three transition lines are expected to meet at a
multicritical point, see Fig.~\ref{phdiagn2}.

The three phases can be characterized using the renormalization-group invariant
quantities $U_{V/T}$ and $R_{V/T}$. In the disordered phase $R_V=R_T=0$,
while $U_V$ and $U_T$ take in the O(N) model the values 
\begin{equation}
U_V = {N+1\over N}, \qquad\qquad U_T = {N^2 + 1\over N^2 - 1}.
\label{U-disorderedphase}
\end{equation}
In the vector-ordered phase $R_V=R_T=\infty$ and $U_V=U_T=1$. In the 
tensor-ordered phase $R_V=0$, $R_T=\infty$ and $U_T=1$. As for $U_V$, 
note that the ordering of $Q$ implies the ordering of the absolute
value of each component $z_{\bm x}^a$, and of the relative phases 
between $z_{\bm x}^a$ and $z_{\bm x}^b$. Only the global 
phase of the field does not show long-range correlations.
Thus, ${\bm z}_{\bm x}$ can be written as $\bar{\psi}_{\bm x} {\bm v}$
where ${\bm v}$ is a constant vector and ${\psi}_{\bm x}$ an uncorrelated
phase. Therefore,
we expect $U_V$ to converge to the value appropriate for a disordered 
 O(2) model, i.e., $U_V = 2$. 

As we discussed in Sec.~\ref{u1gb}, for $\gamma = 0$, the model 
we consider should be equivalent to a model with Hamiltonian
\begin{equation}
H = H_z + \tilde{w} \sum_{{\bm x},\mu} \hbox{Re}\, 
\bar{\bm z}_{\bm x} \cdot {\bm z}_{{\bm x} + \hat{\mu}}, 
\label{Hvector}
\end{equation}
i.e., an equivalent breaking of the gauge invariance is obtained by adding 
a ferromagnetic O($2N$)-invariant vector interaction. On the basis of 
an analysis analogous to that presented above, we expect a phase diagram,
in terms of $J$ and $\tilde{w}$,
similar to the one presented in Fig.~\ref{phdiagn2}. The only difference 
should be the behavior of the DV line, that should connect the multicritical
point with the point $J=0$, $\tilde{w} = w_c$, where $w_c$ is the O($2N$)
critical point. By using the relation between the original 
model and the formulation with Hamiltonian (\ref{Hvector}), one 
can understand 
why the gauge fields $\lambda_{{\bm x},\mu}$ can only be critical at transitions
where the vector field has long-range correlations. For $\gamma = 0$, 
zero-momentum correlations of $\hbox{Re}\, \lambda_{{\bm x},\mu}$ can be 
directly related to vector energy correlations in the model with 
Hamiltonian (\ref{Hvector}). Thus, gauge fields become critical
only on the DV and TV lines. 

\subsection{The phase diagram in the presence of an axial gauge fixing}
\label{sec4.F}

Let us finally discuss the expected phase diagram in the presence of an 
axial gauge fixing. The behavior for $w=0$, i.e., in the gauge-invariant
theory, is mostly unchanged. Gauge-invariant quantities are indeed identical 
in the two cases. Vector correlations vary, but are still short-ranged. 
Indeed, the gauge fixing introduces a
ferromagnetic interaction, which is, however, effectively one-dimensional 
(this is evident in the formulation with Hamiltonian (\ref{Haxial-eff})) 
and therefore unable to give rise to long-range correlations.

The behavior for $\gamma\to \infty$ and $J\to \infty$ instead changes.
In this limit, since no gauge degrees of freedom are present, 
the gauge fields converge to
1. Therefore, only two phases are present---a disordered phase and a
vector-ordered phase---and a single DV transition line.
Given these results, it would be possible for the system to have, for finite 
values of $\gamma$ and $J$, a phase diagram without the tensor-ordered 
phase. The numerical data we will show, instead, indicate that the 
qualitative behavior is not changed by the gauge fixing: the phase diagram
is still the one presented in Fig.~\ref{phdiagn2}. However, consistency 
with the limiting cases gives constraints on the large-$J$, large-$\gamma$
behavior of the TV line. If the TV line is given by the equation 
$w = f_{TV}(J,\gamma)$, one should have $f_{TV}(J,\gamma)\to 0$ for 
$J\to\infty$ at fixed $\gamma$ and for 
$\gamma\to\infty$ at fixed $J$. The gauge fixing only shrinks 
the size of the tensor-ordered phase.

\section{Numerical results}
\label{numres}

In this section we discuss our numerical Monte Carlo results for $N=2$ 
for the model with Hamiltonian (\ref{Hext}), focusing mostly on the behavior 
for $\gamma=0$.
They provide strong evidence in support of the phase diagram 
sketched in Fig.~\ref{phdiagn2} and of the theoretical analysis of 
Sec.~\ref{phdiacrbeh}. An
accurate study of the nature of the multicritical 
point deserves further investigations, that we
leave for future work.

The Monte Carlo data are generated by combining Metropolis updates of
the scalar and gauge fields with microcanonical updates of the scalar field.
The latter are obtained by generalizing the usual reflection moves used in 
O($N$) models. Trial states for the Metropolis updates are
generated so that approximately  $30\%$ of the proposed updates are accepted.
Each data point corresponds to a few milion updates, where 
a single update consists of 1 Metropolis sweep and 5 microcanonical
sweeps of the whole lattice. Errors are estimated by using standard jackknife
and blocking procedures. 

We consider cubic lattices of linear size $L$. In the absence of gauge 
fixing we use periodic boundary conditions along all lattice directions. 
We consider $C^*$ boundary conditions ~\cite{KW-91,LPRT-16,BPV-21-gsb}
in simulations in which the axial 
gauge---$\lambda_{{\bm x},3} = 1$ on all sites---is used.

\subsection{The extended model for $N=2$}

\begin{figure}[tbp]
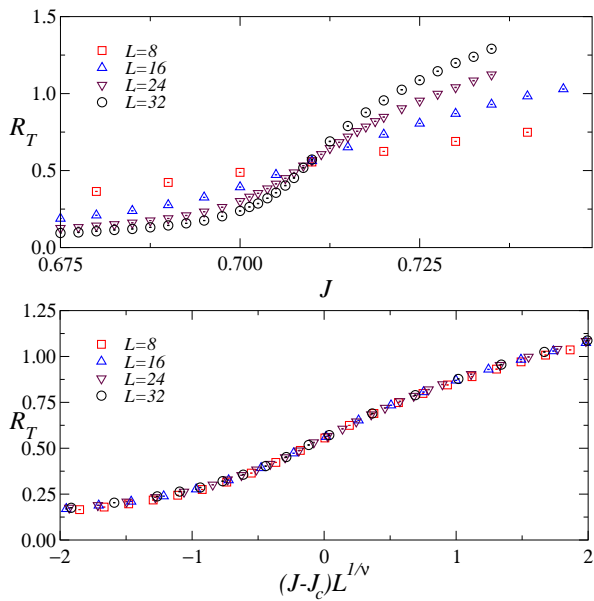

  \includegraphics*[width=0.9\columnwidth]{RTJ_nccp1_kappa0p0_msq0p3.eps}
  \includegraphics*[width=0.9\columnwidth]{RTJ_scal_nccp1_kappa0p0_msq0p3.eps}
  \caption{Numerical data for $\gamma=0$ and $w=0.3$. Top: 
    tensor correlation-length ratio  $R_T$
    versus $J$. Bottom: $R_T$ versus $(J-J_c)L^{1/\nu}$, 
    with $J_c=0.7097$ and the O(3) critical
    exponent $\nu \approx 0.7117$. }
\label{r0p3data_1}
\end{figure}

\begin{figure}[tbp]
  \includegraphics*[width=0.9\columnwidth]{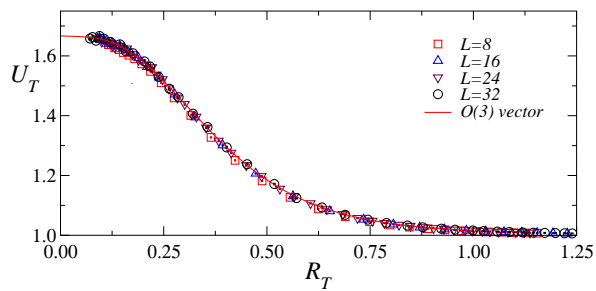}
  \caption{Numerical data for $\gamma=0$ and $w=0.3$. We report
    $U_T$ versus $R_T$ and 
    the universal scaling curve $U_T(R_T)$, defined in~Eq.~(\ref{yfur}), 
    for the O(3) vector model (Eq.~(\ref{FO3V}) in the Appendix).}
\label{r0p3data_2}
\end{figure}

\begin{figure}[tbp]
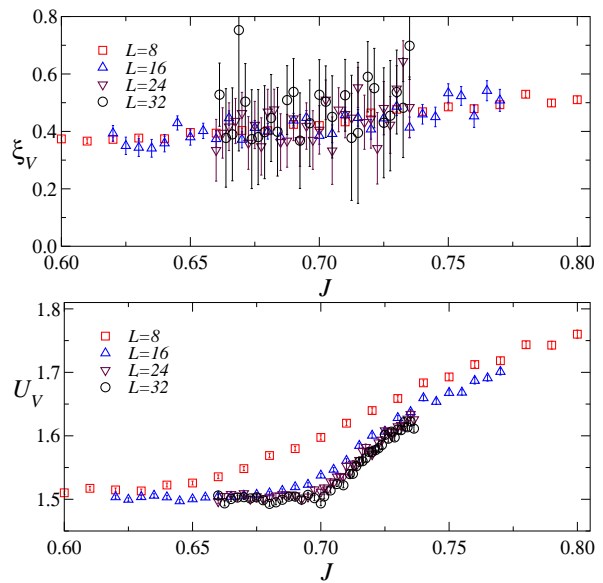

  \includegraphics*[width=0.9\columnwidth]{xiVJ_nccp1_kappa0p0_msq0p3.eps}
  \includegraphics*[width=0.9\columnwidth]{UVJ_nccp1_kappa0p0_msq0p3.eps}
  \caption{Numerical data for $\gamma=0$ and $w=0.3$.
    Results for  $\xi_V$ (top) and $U_V$ (bottom) as a function of $J$
    across the transition, which show that the vector degrees of freedom
    are disordered in the whole critical region.}
\label{r0p3data_3}
\end{figure}

\begin{figure}[tbp]
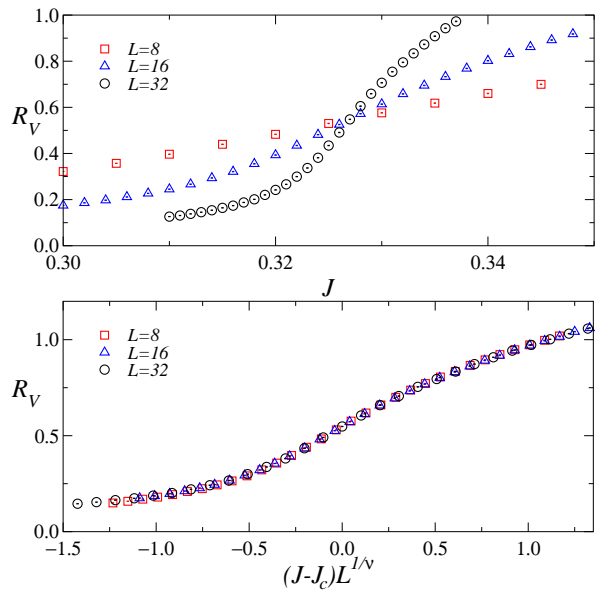

  \includegraphics*[width=0.9\columnwidth]{RVJ_nccp1_kappa0p0_msq2p25.eps}
  \includegraphics*[width=0.9\columnwidth]{RVJ_scal_nccp1_kappa0p0_msq2p25.eps}
  \caption{Numerical data for $\gamma=0$ and $w=2.25$.
   Top: Vector correlation-length ratio $R_V$ versus $J$.
   Bottom: $R_V$ versus $(J-J_c) L^{1/\nu}$, using $J_c=0.3270$ and 
   the O(4) critical exponent $\nu = 0.750$. }
\label{r2p25data_1}
\end{figure}

We start the numerical investigation of the phase diagram of the model 
with Hamiltonian (\ref{Hext}), by studying the critical behavior in the 
small-$w$ region, where we expect transitions to belong to the DT line. 
Since for $J=\infty$ the tensor-ordered phase corresponds to
$w < w_c \approx 0.454$, we perform simulations keeping $w = 0.3 < w_c$ 
fixed and varying $J$. We observe criticality in the 
tensor channel for $J\approx 0.71$, see the upper panel in 
Fig.~\ref{r0p3data_1}. The data for $R_T$ are fully consistent with an 
O(3) critical behavior, as it is evident from the lower panel where we show 
a scaling plot using the O(3) critical exponent $\nu$ and the estimate
$J_c = 0.7097(1)$ of the critical point. Stronger evidence for O(3) behavior 
is provided in Fig.~\ref{r0p3data_2}, where data for
$U_T$ are reported as a function of $R_T$ and compared with the universal
curve of the O(3) vector model, obtaining excellent agreement.

Finally, we check that vector degrees of freedom are not critical at the
transition: for all the values of $J$ studied, $\xi_V$ is very small, see
Fig.~\ref{r0p3data_3}, and the same is true for the susceptibility
$\chi_V$, which takes the value $\chi_V\approx 1.9$ in the transition region
(not shown); note that $\chi_V=1$ for every $J$ when $w=0$. The 
vector Binder parameter
$U_V$ is also reported in Fig.~\ref{r0p3data_3}. As expected, it is 
approximately equal to 
3/2 in the disordered phase, see Eq.~(\ref{U-disorderedphase}),
and increases towards to XY value 2 as J increases.

We next investigate the behavior for large $w$ values, where we expect the 
DV line. Vector and tensor 
correlations should simultaneously order, displaying O(4) vector critical 
behavior.
Again we perform simulations at fixed $w$, choosing $w=2.25$.

\begin{figure}[tbp]
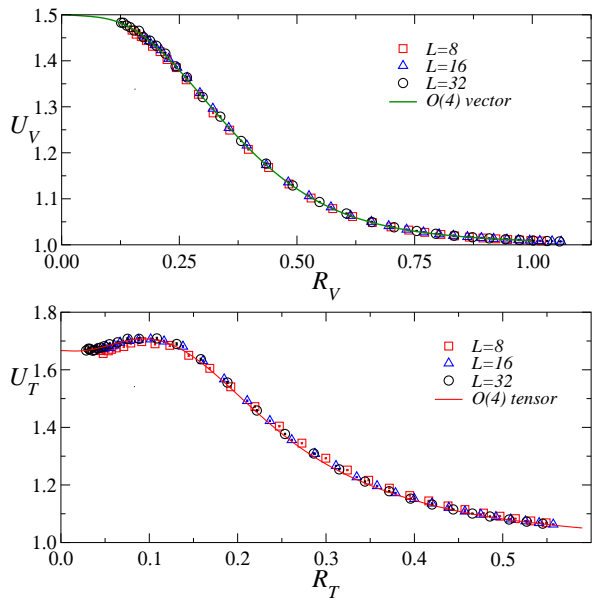

  \includegraphics*[width=0.9\columnwidth]{UVRV_nccp1_kappa0p0_msq2p25.eps}
  \includegraphics*[width=0.9\columnwidth]{UTRT_nccp1_kappa0p0_msq2p25.eps}
  \caption{Numerical data for $\gamma=0$ and $w=2.25$.
    Plots of $U_V$ versus $R_V$ (top) and of $U_T$ versus $R_T$ (bottom).
    The numerical data are compared with the 
    universal scaling curves $U_V(R_V)$ (top) and $U_T(R_T)$ (bottom),
    defined in Eq.~(\ref{yfur}), computed in the O(4)
    vector model (Eqs.~(\ref{FO4V}) and (\ref{FO4T}) in the  Appendix).  
    }
\label{r2p25data_2} 
\end{figure}

In Fig.~\ref{r2p25data_1} we show $R_V$ as a function of $J$: a transition 
is identified for $J=J_c= 0.3270(1)$. The data show very good scaling 
when plotted against $(J-J_c) L^{1/\nu}$, using the O(4) exponent 
$\nu = 0.750$. The O(4) nature of the transition is further confirmed
by the plots of $U_V$ and $U_T$ versus $R_V$ and $R_T$, respectively,
shown in Fig.~\ref{r2p25data_2}. The numerical data fall on top of the
scaling curves $U_V(R_V)$ and $U_T(R_T)$ computed in the O(4) vector model.

\begin{figure}[tbp]
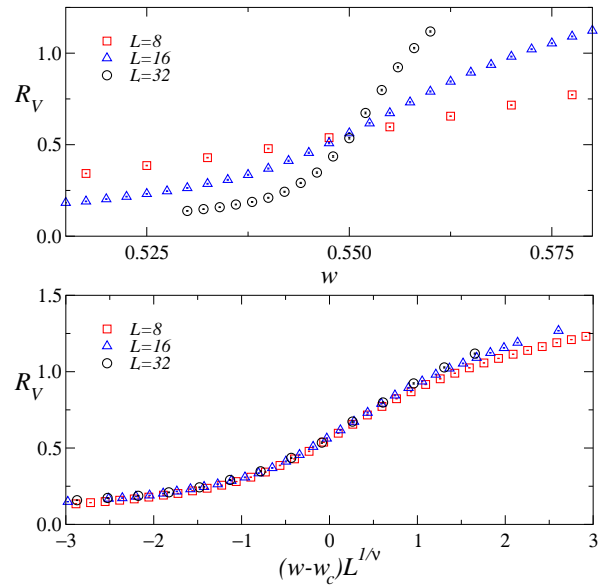

  \includegraphics*[width=0.9\columnwidth]{RVmsq_nccp1_kappa0p0_J1p0.eps}
  \includegraphics*[width=0.9\columnwidth]{RVmsq_scal_nccp1_kappa0p0_J1p0.eps}
  \caption{
    Numerical data for $\gamma=0$ and $J=1$.  Top: Vector correlation-length
    ratio $R_V$ versus $w$. Bottom:  $R_V$ versus $(w-w_c) L^{1/\nu}$,
    using the O(2) value $\nu = 0.6717$. The critical point is located 
    at $w_c = 0.5505(3)$.} 
\label{J1p0data_1}
\end{figure}

Finally, we performed a set of simulations, to investigate the nature of the 
TV line, that separates the two phases in which the tensor degrees of 
freedom are ordered. As discussed in Sec.~\ref{sec4.E},  the TV line 
can be identified using vector observables that should display 
O(2) critical behavior.
Given that the DT line ends at $J_c = 0.7102$, $w = 0$, 
we fixed $J=1$ and increased $w$. As evident from Fig.~\ref{phdiagn2}, 
this choice should allow us to observe the TV line.

\begin{figure}[tbp]
  \includegraphics*[width=0.9\columnwidth]{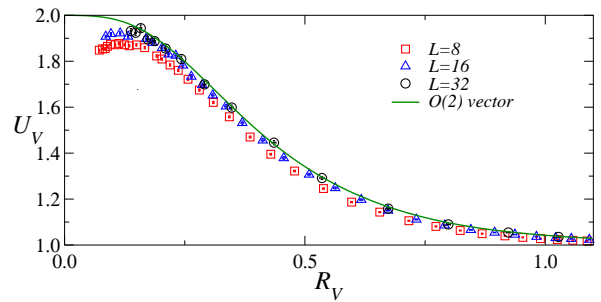}
  \caption{Numerical data for $\gamma=0$ and $J=1$. 
    We report $U_V$ versus $R_V$ and
    the universal scaling curve $U_V(R_V)$, defined in~Eq.~(\ref{yfur}),
    for the O(2) vector model (Eq.~(\ref{FO2V}) in the Appendix).}
\label{J1p0data_2}
\end{figure}

\begin{figure}[tbp]
  \includegraphics*[width=0.9\columnwidth]{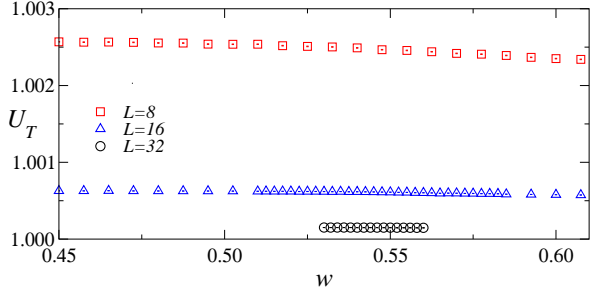}
  \caption{Numerical data for $\gamma=0$ and $J=1$.
    Estimates of $U_T$ versus $w$ across the transition: 
    the tensor degrees of freedom are ordered on both sides of the 
    transition [$w_c = 0.5505(3)$].}
\label{J1p0data_3}
\end{figure}

In Fig.~\ref{J1p0data_1}, we report $R_V$ versus $w$. A crossing point 
is detected for $w_c \approx  0.5505$. Close to it, vector data are 
fully consistent
with an O(2) critical behavior. This is further confirmed by the results 
reported in Fig.~\ref{J1p0data_2}. The data of the vector Binder parameter,
when plotted versus $R_V$, are fully consistent with the corresponding universal
curve computed in the vector O(2) model.
As a further check that the
transition belongs to the TV line, in Fig.~\ref{J1p0data_3} we report $U_T$
versus $J$: 
$U_T$ converges to 1 by increasing the size of the lattice on both 
sides of the transition, confirming that tensor modes are fully magnetized.

\subsection{Simulations in the presence of an axial gauge fixing}

\begin{figure}[tbp]
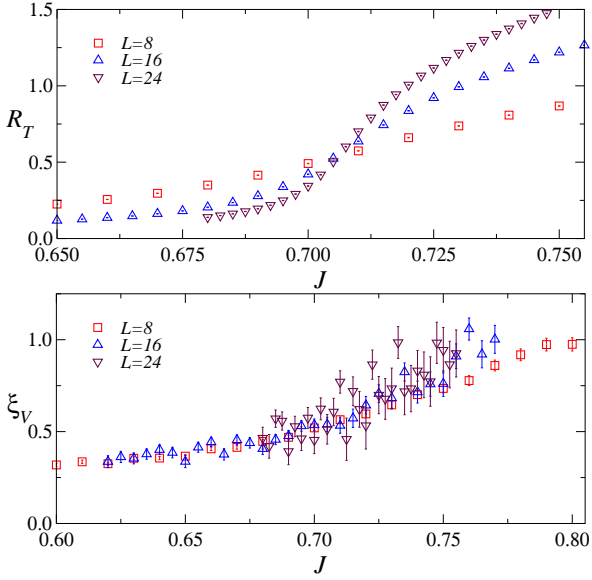

  \includegraphics*[width=0.9\columnwidth]{RTJ_TGF_nccp1_kappa0p0_msq0p1.eps}
  \includegraphics*[width=0.9\columnwidth]{xiVJ_TGF_nccp1_kappa0p0_msq0p1.eps}
  \caption{
Numerical data for $\gamma=0$ and $w=0.1$, in the axial gauge
with $C^*$ boundary conditions. Behavior of $R_T$ (top) and $\xi_V$
(bottom) across the transition [$J_c = 0.706(1)$]. 
While a crossing point for $R_T$ is clearly seen, 
vector degrees of freedom are disordered for all values of $J$.}
\label{axial_r0p2data_1}
\end{figure}

\begin{figure}[tbp]
  \includegraphics*[width=0.9\columnwidth]{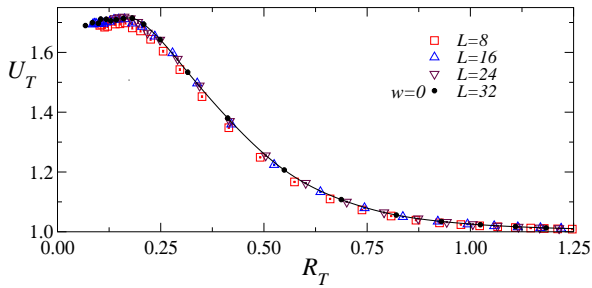}
  \caption{Numerical data for $\gamma=0$ and $w=0.1$, in the axial gauge
    with $C^*$ boundary conditions. The estimates of 
    $U_T$ versus $R_T$ obtained for $w=0.1$ (results for $L=8,16,24$)
    are compared with results for the gauge-invariant model with 
    $w=0$ (results for $L=32$).
    The continuous line is a spline interpolation of the $w=0$ data.
    The consistency of the data signals 
    that the two transitions belong to the same O(3) universality class.}
\label{axial_r0p2data_2}
\end{figure}

We now discuss the extended model with Hamiltonian (\ref{Hext}) in the presence
of the axial gauge fixing. We use C$^*$ boundary conditions
\cite{KW-91,LPRT-16,BPV-21-gsb}, so that we can require $\lambda_{{\bm x},3} =
1$ on all sites.  The purpose of the simulations is that of understanding if 
a tensor-ordered phase as well as a DT transition line are present,
so that gauge invariance is recovered in the critical limit for 
finite small values of $w$. As we shall see,
the answer is positive: the gauge fixing does not change the qualitative 
shape of the phase diagram.

As we have discussed in Sec.~\ref{sec4.F},  even if the qualitative phase 
diagram is unchanged, the DT phase should shrink and the TV line should 
get closer to the $w=0$ axis.
For this reason, we decided to perform simulations at fixed $w = 0.1$. 
The vector and tensor correlation lengths are reported in
Fig.~\ref{axial_r0p2data_1} as a function of $J$. The data for $R_T$ 
have a crossing point at $J_c = 0.706(1)$, which is very close 
to the critical point for $w = 0$, $J_c = 0.7102(1)$ \cite{PV-19-AH3d}.
The tensor degrees of freedom are critical at the transition.
The vector ones are instead disordered and $\xi_V$ is of 
order 1 across the transition. Thus, the data confirm the existence of a DT line
also in the presence of the axial gauge fixing. As an additional check, in 
Fig.~\ref{axial_r0p2data_2} we plot $U_T$ against $R_T$. The data are compared 
with the results for the gauge-invariant model ($w=0$) with the same C$^*$ 
boundary conditions (we consider results with $L=32$, which should 
provide a good approximation of the asymptotic curve). The agreement is 
excellent, confirming that the transition belongs indeed to the DT line,
where O(3) behavior is expected. 
As a side remark, note that the 
O(3) curves reported in Figs.~\ref{axial_r0p2data_2} 
and~\ref{r0p3data_2} are different, 
since the scaling curve depends on the boundary conditions. 

We also performed simulation with $w=0.2$. In this case, we observe two 
very close transitions, that are naturally identified with transitions on the 
DT and TV line. They provide an approximate estimate of the 
multicritical point, $w_{mc} \approx 0.2$, $0.69\lesssim J_{mc} \lesssim 0.70$.
Note that the size of the tensor-ordered phase is, not surprisingly,
significantly smaller
than in the absence of gauge fixing. Indeed, in the latter case $w_{mc} \gtrsim 
w_c(J=1) \approx 0.55$.

\subsection{Finite $\gamma$ results for the pure gauge model}

\begin{figure}[tbp]
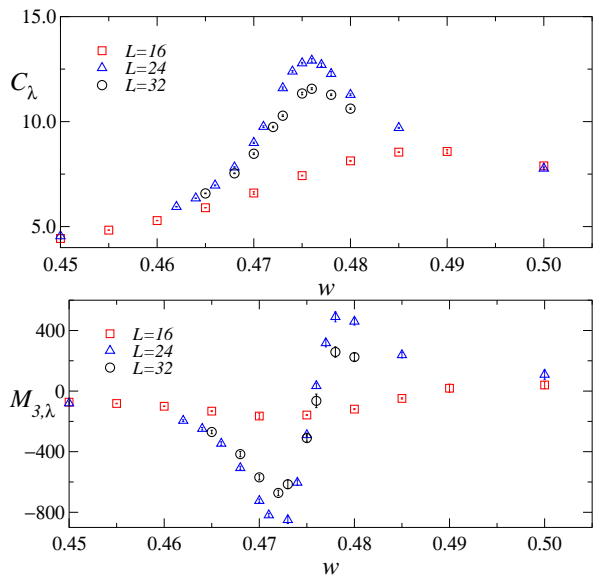

  \includegraphics*[width=0.9\columnwidth]{Clambda_nccp1_kappa2p5_J1p0.eps}
  \includegraphics*[width=0.9\columnwidth]{M3lambda_nccp1_kappa2p5_J1p0.eps}
  \caption{Numerical data for $\gamma=2.5$ and $J=0$ (pure gauge model).
    Top panel: specific heat 
    $C_\lambda$; Bottom panel: third moment $M_{3,\lambda}$.}
\label{Puregauge}
\end{figure}

At $\gamma=\infty$, the phase diagram is characterized by four 
transition lines, see Sec.~\ref{sec4.C}. In addition to the lines 
reported in Fig.~\ref{phdiagn2}, there is a line where $\lambda_{{\bm x},\mu}$
is critical and which separates two phases with no tensor or vector order. 
As we have discussed in Sec.~\ref{sec4.E}, such a line is not expected to 
occur at $\gamma =0$. We wish now to provide evidence that such a line 
does not exist for any finite $\gamma$. As this line starts on the 
$J=0$ line, we consider the pure gauge model with Hamiltonian
\begin{equation}
H = H_\lambda(\lambda) + H_b(\lambda).
\end{equation}
For $\gamma=\infty$, there is an XY transition for $w=w_{c,XY}\approx 0.454$.
We wish now to verify whether there is a 
transition for large, but finite values 
of $\gamma$. If it were present, it would imply the presence of a fourth
transition line in the phase diagram. For this purpose, we have studied 
the model for $\gamma = 2.5$. For this value of $\gamma$ the gauge fields 
are significantly ordered and indeed $\langle \Pi_{{\bm x},\mu\nu}\rangle
\approx 0.93$ 
in the relevant region $w\approx w_{c,XY}$. To detect the transition, we have 
considered cumulants of the energy
\begin{eqnarray}
C_\lambda &=& {1\over V w^2} \langle (H_b - \langle H_b\rangle)^2 \rangle  ,
\nonumber \\
M_{3,\lambda} &=& {1\over V w^3} \langle (H_b - \langle H_b\rangle)^3 \rangle
.
\end{eqnarray}
At an XY transition, the specific heat has a nondiverging maximum
while the third cumulant should have 
a positive and a negative peak \cite{SSNHS-03}, 
both diverging as $L^{3/\nu-3} = L^{(1+\alpha)/\nu} \approx L^{1.47}$. 
In Fig.~\ref{Puregauge}, we show the two quantities as a function of $w$.
The specific heat has a maximum at $w \approx 0.476$, which might in principle 
signal an XY transition. However, $M_{3,\lambda}$ is not diverging. It
increases significantly as $L$ changes from 16 to 24, but its maxima decrease 
as $L$ varies from 24 to 32. We are evidently observing a crossover behavior, 
due to the transition present for infinite $\gamma$. Simulations 
do not provide evidence of finite-$\gamma$ transitions and therefore, of a 
fourth transition line in the phase diagram.

\section{Conclusions}
\label{conclu}

In this work we discuss the role of GSB perturbations in lattice gauge 
models. In Ref.~\cite{BPV-21-gsb} we studied the AH model with 
noncompact gauge fields, analyzing the role of GSB perturbations at 
transitions associated with a charged FP, i.e., where both scalar-matter
and gauge correlations are critical. In that case we showed that there 
is an extreme sensitivity of the system to GSB
perturbations, such as the photon mass in the AH field theory. 
In this work, instead, we discuss the behavior at transitions where 
gauge correlations are not critical and the gauge symmetry has only the role 
of reducing the scalar degrees of freedom that display critical behavior. A
paradigmatic model in which this type of behavior is observed is the 
3D lattice CP$^1$ model or, more generally, 
the lattice AH model with two-component complex scalar
matter and U(1) compact gauge fields. We find that in this case, 
the critical gauge-invariant modes
are robust against GSB perturbations: for small values of the GSB coupling,
the critical behavior (continuum limit) is the same as in the 
gauge-invariant model.

Note that, even when gauge symmetry is explicitly broken, the existence of a
gauge-symmetric limit is very useful to understand the physics of the model. If
indeed one directly studies the gauge-broken model, a detailed analysis of the
nonperturbative dynamics of the model would be required to identify the nature
of the critical degrees of freedom, which instead emerges naturally in the
gauge-invariant limit. Looking for gauge-invariant limits or deformations 
might thus be a useful general strategy to pursue in order to understand 
the origin of ``unusual'' orderings 
and transitions (the DT and the TV lines in the present model 
or in the models with Hamiltonians (\ref{hlaeffexpmix}) and 
(\ref{Haxial-eff})). 

In our study we consider the $N=2$ lattice AH model with Hamiltonian 
(\ref{Hdef}). The model has two phases \cite{PV-19-AH3d} separated by 
a continuous transition line driven by the
condensation of the gauge-invariant operator $Q$
defined in Eq.~(\ref{qdef}). The nature of the transition is independent of the 
gauge coupling and the is same as that in the CP$^1$ model, which, in turn, is 
the same as that in the O(3) vector model.

We add to the Hamiltonian (\ref{Hdef}) the GSB term
$H_b=- w \sum_{{\bm x},\mu} {\rm Re} \,\lambda_{{\bm
    x},\mu}$, where 
the parameter $w$ quantifies the strength of the perturbation.
A detailed numerical MC study, supported by the analysis of some limiting
cases,
allows us to determine the phase diagram of the model, see Fig.~\ref{phdiagn2}.
Its main features are summarized as follows.

\begin{itemize}

\item
The parameter $w$ is an irrelevant RG perturbation of the gauge-invariant
critical behavior (continuum limit). 
Therefore, even in the presence of a
finite small GSB perturbation, the critical behavior is the same as that 
of the gauge-invariant  CP$^1$ universality class.
The gauge-breaking effects of the linear perturbation disappear in 
the large-distance behavior.

\item
As sketched in Fig.~\ref{phdiagn2}, the phase diagram of the model with
Hamiltonian Eq.~(\ref{Hext}) presents three phases for $\gamma=0$:
a disordered phase for small $J$ and two 
ordered phases for large $J$.  There is a
tensor-ordered phase for small values of $w$, in which the operator $Q$
defined in Eq.~(\ref{qdef}) condenses, while the vector correlation
$G_V$ defined in Eq.~(\ref{vectorfu}) is short ranged.  For
sufficiently large values of $w$, the ordered phase is
characterized by the condensation of
the vector matter field.  

The three phases are separated by three different transition lines that
presumably meet at a multicritical point. It would be interesting to 
understand its nature, but we 
have not pursued this point further.

Most of our results are obtained for the case $\gamma=0$, i.e., for the 
CP$^1$ model, but we expect the same phase diagram also for any finite $\gamma$,
with strong crossover effects for large values of $\gamma$.

\item
Transitions along the DT line, that separates the disordered
phase from the tensor-ordered phase, belong to the O(3) vector universality
class, as the transition for $w=0$.

\item
Transitions along the DV line, that separates the disordered
phase from the vector-ordered phase, belong to the 
O(4) vector universality class \cite{BPV-21-gsb}. Note the 
effective symmetry enlargement at the transition---from U(2) to O(4)---that
was predicted using RG arguments based on the correspoding LGW
theory~\cite{BPV-21-gsb}.
Of course, the symmetry enlargement is limited to the critical regime.
Outside the critical region,
one can only observe the global U(2) symmetry of the model.

\item
Transitions along the TV line, that separates 
the tensor and the vector ordered phases, are associated with the condensation
of the global phase of the scalar field, which is disordered in the 
tensor-ordered phase and ordered in the vector-ordered phase.
Continuous transitions along the TV line belong to the O(2) or XY
universality class.  Note that this mechanism realizes an unusual
phenomenon: two ordered phases are separated by a continuous
transition line.

\item
We have also studied the model in the presence of a gauge fixing. 
We considered the axial gauge fixing defined in Eq.~(\ref{axialgauge}).
We find the phase diagram to be  qualitatively similar
to that found in the absence of a gauge fixing, see Fig.~\ref{phdiagn2}. 
In particular, for sufficiently small $w$, the critical behavior is the 
same as in the gauge-invariant theory.

\end{itemize}

We have not analyzed the behavior for $N>2$. Also in this case, we expect 
a phase diagram with three different phases, as for $N=2$. However, the 
nature of the transition lines might be different. 
Since the transitions in the gauge-invariant lattice CP$^{N-1}$ models
are of first-order for any $N>2$, the transitions along the DT line should be
of first order. Transitions along the TV line should still belong to the 
XY universality class, if they are continuous, while transitions along
the DV line are expected to belong
to the O($2N$) vector universality class with the effective
enlargement of the symmetry of the critical modes from U($N$) to
O($2N$).

There are several interesting extension of our work. One might
consider GSB terms involving higher powers of the link 
variables, for instance 
$H_{b,q}=-w\sum_{{\bm x},\mu} {\rm Re}\,\lambda_{{\bm x},\mu}^q$, 
which leave a residual discrete ${\mathbb Z}_{q}$
gauge symmetry. In this case we may have a more complex phase diagram
where the residual gauge symmetry may play a role. One may also 
consider the same linear perturbation in the AH model with higher-charge
compact matter fields \cite{BPV-20-q2}, which has a more complex 
phase diagram than the AH lattice model we consider here.

Finally, we mention that several nonabelian
gauge models with multiflavor scalar
matter~\cite{BPV-19-sqcd,BPV-20-son,BFPV-21-adj} have transitions
where gauge correlations do not become critical. Gauge symmetry is 
only relevant for defining the critical modes
and therefore the symmetry-breaking pattern associated with the 
transition, as in the lattice AH model we have considered here.
For this class of nonabelian models, one might investigate 
the role of similar GSB terms, for instance of a perturbation
$H_b = - w
\sum_{{\bm x},\mu} \hbox{Re Tr} \, U_{{\bm x},\mu}$ where $U_{{\bm
    x},\mu}$ are the gauge variables associated with the lattice links 
\cite{Wilson-74}, to understand whether a gauge-invariant continuum limit
is still obtained for small values of $w$.

\acknowledgments

Numerical simulations have been performed on
the CSN4 cluster of the Scientific Computing Center at INFN-PISA.

\appendix
\section{Universal scaling curve in O($N$) vector models}

In this Appendix we collect the expressions of the scaling curves 
$F_U(R)$, defined in Eq.~(\ref{yfur}), for periodic boundary conditions 
and cubic lattices $L^3$, computed in the O(2), O(3), and O(4) vector models.
We report here some simple parametrizations.
The error on these expressions should be less than 0.5\%.

On the TV line, the vector data have been compared with analogous vector 
data computed in the XY model. The XY curve $U_V=F_U(R_V)$ is given by
\begin{eqnarray}
&& F_{U}(x) = 2 \nonumber  \\
&& \qquad + 27.508562 x^2 - 216.397337 x^3 + 360.327374 x^4 \nonumber \\
&& \qquad - 307.205086 x^5 + 133.83076 x^6 - 23.718357 x^7 \nonumber \\
&& \qquad - (1 - e^{-7 x^2}) (4.038703 - 5.785571 x) \nonumber \\
&& \qquad + 22.958929 x^2 (1 - e^{-8x}) , 
\label{FO2V}
\end{eqnarray}
valid for $x < 1.1$. 

Along the DT line, the scaling curve $U_T=F_U(R_T)$ in the tensor sector
is the same as the scaling curve in the O(3) model for vector quantities 
(i.e., computed using correlations of $m^a = \sum_x s^a_{\bm x}$, 
where $s^a_{\bm x}$ is a three-dimensional unit spin):
\begin{eqnarray}
&& F_{U}(x) = {5\over 3}  
- 47.83889 x^2  + 58.48967 x^3  - 67.02068 x^4   \nonumber \\
&& \qquad + 38.408855 x^5 - 8.8557348 x^6  \nonumber \\
&& \qquad + x (3.0263535 + 23.139470 x) (1- e^{-15x}),
\label{FO3V}
\end{eqnarray}
valid for $x < 1.0$. 

Finally, the scaling curves $U_V=F_U(R_V)$ and $U_T=F_U(R_T)$ along the DV line
can be computed in the O(4) model.  The vector curve $U_V=F_U(R_V)$
corresponds to the vector curve in the O(4) model, 
which can be parametrized as 
\begin{eqnarray}
&& F_{U}(x) =  {3\over 2} -
48.804243 x^2 + 64.371024  x^3 \nonumber \\
&& \qquad - 85.177814  x^4 + 62.735307 x^5  \nonumber \\ 
&& \qquad - 24.558342  x^6 + 3.998991  x^7 \nonumber \\
&& \quad + x (3.039763 + 23.904433 x) (1 - e^{-15x}),
\label{FO4V}
\end{eqnarray}
valid for $x < 1.0$. The tensor curve
$U_T=F_U(R_T)$ for the tensor operator $Q$ 
can also be related to an O(4) scaling curve. The relation 
is discussed in App. B of Ref.~\cite{PV-19-AH3d}. We parametrize the 
O(4) results as 
\begin{eqnarray}
&& F_{U}(x) =  {5\over3} - 2077.568536 x^2 + 6198.556568 x^3 \nonumber \\
&& \quad - 15494.694968 x^4 + 22321.613611 x^5 \nonumber \\
&& \quad -17453.201432 x^6 + 5754.439605 x^7 \nonumber \\
&& + x (128.626365 + 713.495184 x) (1 - e^{-16 x}),
\label{FO4T}
\end{eqnarray}
valid for $x < 0.6$.

\end{document}